# Utility Constrained Energy Minimization In Aloha Networks

Amirmahdi Khodaian, Babak H. Khalaj, Mohammad S. Talebi
Electrical Engineering Department
Sharif University of Technology
Tehran, Iran
khodaian@ee.shrif.edu, khalaj@sharif.edu, mstalebi@ee.sharif.edu

***Abstract*—In this paper we consider the issue of energy efficiency in random access networks and show that optimizing transmission probabilities of nodes can enhance network performance in terms of energy consumption and fairness. First, we propose a heuristic power control method that improves throughput, and then we model the Utility Constrained Energy Minimization (UCEM) problem in which the utility constraint takes into account single and multi node performance. UCEM is modeled as a convex optimization problem and Sequential Quadratic Programming (SQP) is used to find optimal transmission probabilities. Numerical results show that our method can achieve fairness, reduce energy consumption and enhance lifetime of such networks.**

## I. INTRODUCTION

Energy efficiency of wireless networks has received considerable attention in recent years. Especially, in sensor networks, nodes are not rechargeable and lifetime of the node is equal to its battery lifetime. It is therefore necessary to avoid any waste of energy and ensure network longevity. On the other hand, protocols of these networks should be simple and distributed and thus, random access protocols are frequently used in such networks. Slotted Aloha [1] is the basic and most studied random access in which nodes transmit their packets with certain probability in every time slot. Usually collision model is used to analyze performance of the protocol [1]. In this model, it is assumed that when two packets arrive at the receiver collision occurs and none of them can be decoded with negligible error. However, this model is pessimistic and provides a lower bound on system performance, since in many cases the packet with largest power can capture the channel and be received correctly.

Several works (e.g. [2], [3], [4]) have studied the capture effect and used power control to intensify it. These schemes usually suppress the weak signals and enhance the strong ones in order to increase throughput of the network. However, it is clear that such an approach will result in unfairness among nodes. For example, an algorithm proposed by Metzner[2] reduces transmitter power of far nodes in order to enhance throughput but will simultaneously diminish success probabilities of such nodes. In contrast, if the power control algorithm tries to achieve fairness among nodes by improving weak signals, capture probability and network throughput will be sacrificed.

General analytic framework for fairness in multi-access wireless channels was first proposed in [5] where it was shown that defining fairness in these channels is equivalent to specifying a utility function and the logarithmic utility was used to provide proportional fairness. If we force the utility function to be greater than a threshold, we can expect a level of fairness among nodes in addition to acceptable network throughput. Fairness has been also addressed in wireless sensor networks in order to ensure that data is collected from different regions of the network [6].

In this paper, we use power control and optimize transmission probabilities in order to minimize energy consumption of the network and provision fairness. Our heuristic power control scheme increases capture probability without affecting fairness, and the computed optimal transmission probabilities guarantee fairness and maximize energy efficiency of the network. Most of the earlier studies on random access have only focused on network throughput and few of them have considered fairness. In addition, to the best knowledge of authors, our work is the first one that minimizes energy while ensuring fairness among network nodes. In [5], [7], [8], and [9], fairness is extensively studied but none of them evaluated energy consumption of nodes. The multi-group model is used in [7] and a retransmission control policy that enhances fairness is suggested, although the optimality of the algorithm was not proven. In [8] and [9], optimal transmission probabilities were found to achieve fairness but only collision model was used and energy consumption was ignored. Energy efficiency of the network was considered in [3], [10], and [11], however, they have mainly investigated throughput-energy tradeoff in Aloha networks without taking fairness into account.

The structure of this paper is as follows. In section II we introduce system model and main assumptions of the work. Our power control and node classification method is described in section III. We formulate our optimization problem as a function of transmission probabilities in section IV and prove its convexity. In section V we propose a method for reducing messages sent from base station to nodes and present the final algorithm. Numerical results and conclusions are given in sections VI and VII, respectively.

This work is supported in part by Iran Telecommunication Research Center (ITRC).

## II. SYSTEM DISCRIPTION

We consider a system in which a finite number of nodes desire to transmit their packets to a *Base Station* (BS) (Fig. 1). The nodes use slotted Aloha in which channel is divided into timeslots with duration $T$ which is equal to the time required to send a packet. As we will discuss in section III, nodes are divided into $M$ groups, there are $n_i$ nodes in group $i$ and node $j$ of group $i$ is denoted by $(i, j)$. Each node transmits in one slot with probability $q_{ij}$ (called transmission probability) and the amount of energy it uses to transmit a packet is $E_{ij}$. Similar to [10], we suppose that transmission and retransmission rates are the same for all nodes. It is also assumed that BS estimates channel gains ($G_{ij}$) from received packets. Thus, the optimization algorithm in receiver incorporates channel state information of the current slot. This implies independency of the algorithm from distribution of nodes and statistical characteristics of channel.

A feedback channel is assumed to exist from BS to nodes, and is used for acknowledgment, synchronization and controlling transmission parameters of nodes. It is assumed that the network changes very slowly, and coherence time of channels is large. Therefore, updating the node parameters does not take place frequently and the effect of this feedback on total energy consumption and throughput is negligible. We have also assumed that when there is no power control, all nodes use the same power, $P$, to transmit a packet. In this work, we do not take stability and delay issues into account and assume that they are controlled by setting appropriate source rates at higher network layers.

## III. CAPTRE EFFECT AND POWER CONTROL

### A. *Perfect Capture Model*

If signal to interference and noise ratio (SINR) of a received packet in slotted Aloha network is above a certain threshold and appropriate coding is used, reliable communication is possible [12]. If we denote interference and noise terms by $I$ and $N$ respectively, and use fixed type of modulation and coding, the packet with power $P$ can be received successfully if:

$$\frac{PG}{I+N} > \beta \tag{1}$$

where G is the channel pathgain and β is SINR threshold. It means that a packet will capture the channel if its SINR is above a specific threshold. Noise term is neglected in our analysis since interference among nodes is the most important factor in multi-access wireless networks. Also, similar to [2], we assume that a packet would be received correctly if and only if it has the strongest power among all of the received packets. This is called *perfect capture model* and is close to SINR threshold model for our system because:

- Our power control (which we will subsequently describe in this section) classifies nodes and ensures that received power from nodes of different classes is quite different.

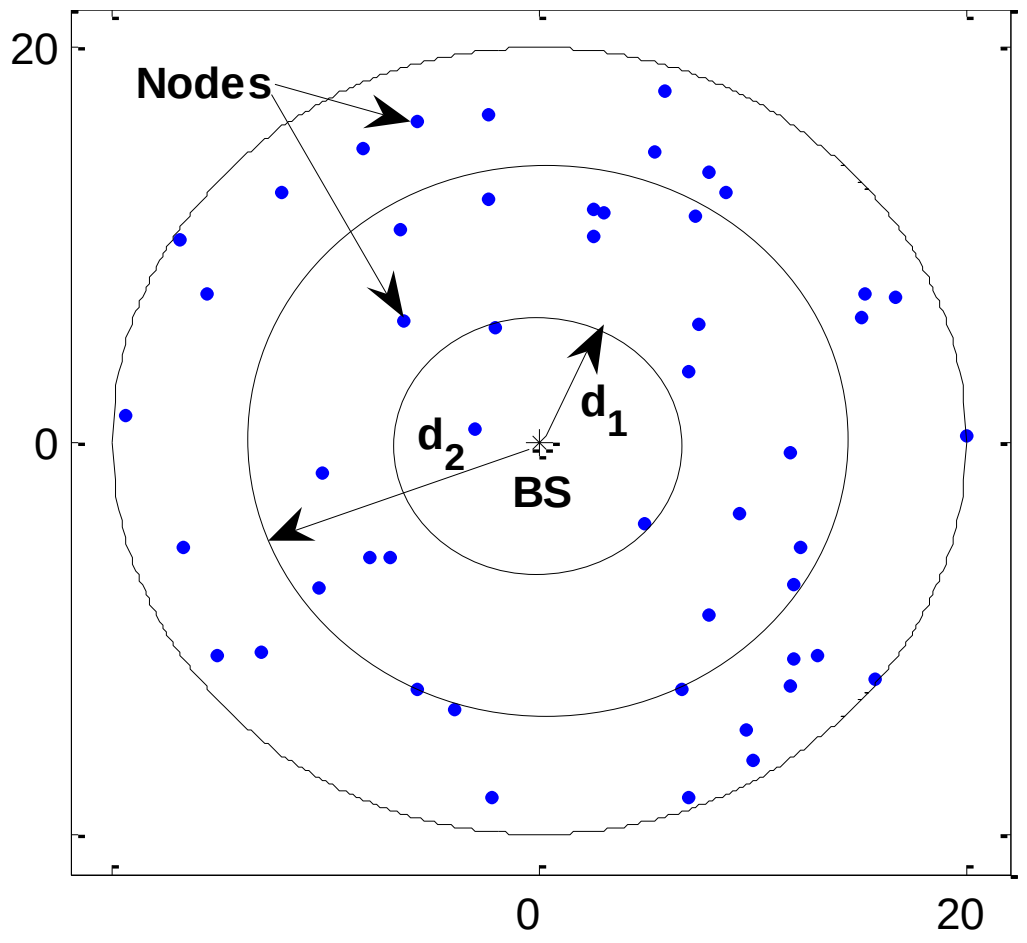

Figure 1. A typical network with 50 nodes and 20m radius

- Since we are looking for an energy efficient system, transmission probabilities should be set small enough in order to avoid unsuccessful transmissions. Therefore, it is less probable that sum of the interferences caused by packets with low power exceeds the strongest packet.

In addition, numerical results of section V confirm this claim and show that performance of our system with perfect capture model is close to the case that we used SINR threshold model. We should insist here that the assumption of perfect capture model is highly related to power control algorithm, and is not necessarily applicable in general.

### B. *Power Control*

Our power control algorithm first classifies nodes into $M$ groups according to their channel gains. To this end, two thresholds are assigned for the channel gains of nodes in each group. For example, the node $(i, j)$ is in group $i$ if its gain $G_{ij}$ satisfies:

$$G_{i+1} < G_{ij} \le G_i \tag{2}$$

Power control algorithm, (3), ensures that packets of nodes in one group are received with the same power at BS.

$$P_{ij} = P\frac{G_i}{G_{ij}} \tag{3}$$

In order to exploit capture effectively, and increase probability of successful transmission for nodes with higher channel gain, we should set proper threshold levels. We set thresholds as (4) to make certain that when two packets are transmitted simultaneously from group $i$ and $k$ ($i<k$) the packet from group $i$ will be received correctly:

$$G_{i+1} = \frac{G_i}{\beta} \tag{4}$$

As an example, suppose that nodes are distributed uniformly in a circle and channel gains depend only on the distance of nodes from BS (Fig. 1). For this network, gain thresholds are equivalent to distance thresholds and group $i$ consists of nodes that their distances to BS satisfies $d_i<d<d_{i+1}$.

## IV. Optimal Transmission Probabilities

Utility Constrained Energy Minimization (UCEM) problem can be formulated as follows:

$$\begin{aligned} &\min \sum_{i=1}^{M}\sum_{j=1}^{n_i} E_{ij} \\ &s.t. \qquad U > U_c \\ &\qquad\qquad 0 \le q_{ij} \le 1 \end{aligned} \tag{5}$$

where $q_{ij}$ and $E_{ij} = P_{ij}\, q_{ij}$ are transmission probability and average energy consumption of the node *(i,j)*, $U_c$ is the threshold for acceptable value of utility function and $n_i$ is number of nodes in group *i*. In order to solve this problem, it is assumed that power of each node is determined earlier by the power control algorithm.

First, we discuss the utility function and find its maximum value $U_{max}$. Similar to some related work [5] and [8], we use the logarithmic utility function. In other words if $x_{ij}$ represents effective rate of node (*i,j*), then utility function of an *M* group network is given by:

$$U = \sum_{i=1}^{M}\sum_{j=1}^{n_i} \log(x_{ij}) \tag{6}$$

Since *log(x)* goes to negative infinity as *x* goes toward zero, a finite value for this utility function makes sure that effective rates of all nodes are above zero. It is also shown in [5] that logarithmic utility results in proportional fairness among nodes.

If we denote throughput of nodes by $S_{ij}$ then:

$$\log(x_{ij}) = \log(L \cdot R_p) + \log(S_{ij}) \tag{7}$$

where *L* is number of bits in a packet and *Rp=1/T* is packet transmission rate. Thus, utility function can be rewritten as:

$$U = U' + N \cdot \log(L \cdot R_p) \tag{8}$$

$$U' = \sum_{i=1}^{M}\sum_{j=1}^{n_i} \log(S_{ij}) \tag{9}$$

Since *N log(L.Rp)* is constant, in the rest of paper we use *U'* instead of *U* in order to either maximize utility or set a threshold for it. According to perfect capture model a packet sent by a node in group *i* is successfully received if and only if no other node in groups 1 to *i* has sent a packet. Therefore, throughput of this node is:

$$S_{ij} = q_{ij}\left[\prod_{l=1}^{i-1}\prod_{k=1}^{n_l}(1-q_{lk}) \cdot \prod_{\substack{k=1\\k\ne j}}^{n_i}(1-q_{ik})\right] \tag{10}$$

and *U'* is given by:

$$U' = \sum_{i=1}^{M}\sum_{j=1}^{n_i}\left[\log\left(\frac{q_{ij}}{1-q_{ij}}\right) + \sum_{l=1}^{i}\sum_{k=1}^{n_l}\log(1-q_{lk})\right] \tag{11}$$

Rearranging terms we have:

$$U' = \sum_{i=1}^{M}\sum_{j=1}^{n_i}\left[\log\left(\frac{q_{ij}}{1-q_{ij}}\right) + \left(\sum_{k=i}^{M} n_k\right)\log(1-q_{ij})\right] \tag{12}$$

If we show internal terms of the above summation as $\varphi_{ij}(.)$ we have:

$$U' = \sum_{i=1}^{M}\sum_{j=1}^{n_i}\varphi_{ij}(q_{ij}) \tag{13}$$

$\varphi_{ij}(.)$ is a concave function due to the fact that (14) is not positive when there is at least one node in group *M*.

$$\frac{\partial^2\varphi_{ij}}{\partial {q_{ij}}^2} = \frac{-\left(1 + {q_{ij}}^2\sum_{k=i}^{M} n_k - 2q_{ij}\right)}{{q_{ij}}^2\left(1-q_{ij}\right)^2} \tag{14}$$

Thus, *U'* is concave because it is the sum of concave functions [13]. This result implies convexity of UCEM problem since the objective of this problem is a linear function of transmission probabilities. There exists a variety of methods to solve convex optimization problems. Among them we have chosen Sequential Quadratic Programming (SQP) [14] for the reason that our variables are bounded and the initial guess is not far from the optimal solution.

## V. Proposed Algorithm

### *A. Reducing Messeges*

According to the schemes presented in sections III and IV, BS can calculate both transmission powers and optimal transmission probabilities of nodes and broadcast these parameters, however, it is important to reduce the overhead of the messages sent by BS because they should be broadcasted every time that network topology changes. Equivalently, reducing messages (or message size) will enable us to use this algorithm in networks that may vary more frequently.

First, we show that power control algorithm can work in a distributed manner throughout the network. If we assume channels are symmetric, nodes can estimate their channel gains with beacons transmitted by BS. Therefore, they can use threshold levels (which are sent to them at system start-up) in order to indicate which group they belong to, and consequently set their power levels. In this case no online feedback is needed from BS.

We can also reduce messages sent by BS for setting transmission probabilities at the cost of making some calculations in nodes. We will show subsequently that if BS only broadcasts the number of nodes in each group and Lagrange multiplier then every node can calculate its unique optimal transmission probability.

The Lagrangian associated with UCEM problem is:

$$L(\underline{q},\lambda) = \lambda U_c + \sum_{i=1}^{M}\sum_{j=1}^{n_i}\left(P_{ij}q_{ij} - \lambda\varphi_{ij}\right) \tag{15}$$

By setting $\nabla_q L = 0$ and using some algebraic manipulation we will get the following quadratic equation which should be solved at each node:

$$q_{ij}{}^{2}P_{ij} - q_{ij}\left(P_{ij} + \lambda\sum_{k=i}^{M} n_k\right) + \lambda = 0 \qquad (16)$$

Since $U$ is concave function of $q$ Lagrangian is strictly concave and thus, it has a unique maximum. It is also clear that (16) has at least one root in (0, 1) because its right hand side (*r.h.s*) has the following property: *r.h.s* $\leq 0$ when $q_{ij}=1$ and *r.h.s* $\geq 0$ when $q_{ij}=0$.

### B. *Algorithm*

In summary, our algorithm that specifies power and transmission probabilities consists of the following steps:

**Step 1.** At the system startup, BS specifies channel gain thresholds according to SINR threshold and sends them to the nodes.

**Step 2.** Both BS and nodes estimate channel gains and if it was different with previous values they classify nodes with respect to channel gain thresholds.

**Step 3.** BS evaluates power values of all nodes and initiate step 4. Every node sets its power.

**Step 4.** BS solves UCEM problem and determines lagrangian multiplier. BS then broadcasts number of nodes in each group and Lagrangian multiplier to the nodes.

**Step 5.** Nodes set their transmission probabilities by solving (16) and choosing the root which is in (0,1).

## VI. Numerical Results

We have applied our algorithm to the sample network illustrated in Fig. 1, and explored energy-utility tradeoff, comparing our method with the case that all nodes have same transmission probabilities (Hereafter we will call the latter uniform policy). Parameters used for our numerical analysis are given in table 1.

Fig. 2 shows effective rate of nodes achieved by our algorithm. In this figure, nodes are sorted according to their distance to BS. It can be seen that numerical results calculated by the perfect capture model are close to the results of SINR model. Our algorithm is also compared with the uniform policy and it is shown that when the same amount of utility is achieved near and far nodes in our algorithm have less throughput differences than near and far nodes of the uniform policy.

In order to analyze the tradeoff between utility and energy, we have solved UCEM problem and calculated the minimum amount of energy consumed for different utility constraint values. According to Fig.3a, energy consumption is very sensitive to $U_c$ and with small variation of this threshold minimum required energy will be reduced to half. Energy efficiency of our algorithm is compared with uniform policy in Fig.3b and it is observed that our algorithm reduces energy consumption by about 10% for all values of $U_c$.

TABLE I. Simulation Parameters

| | |
|---|---|
| Packet Length ($L$) | 1000 bit |
| Time Slot ($T$) | $5^{msec}$ |
| SINR Threshold ($\beta$) | $6^{dB}$ |
| Channel Gain [a] ($G_{ij}$) | $20 \cdot d_{ij}^{-4}$ |
| Node initial power ($P$) | $200^{mWatt}$ |
| Network Radius ($R$) | $20^{m}$ |
| Number of nodes($N$) | 50 |
| Battery Energy ($E_B$) | $1000^{Jouls}$ |
| Utility Constraint ($U_c$) | 219 |

a. For simplicity we have assumed that channel gains depend only on the distance to BS.

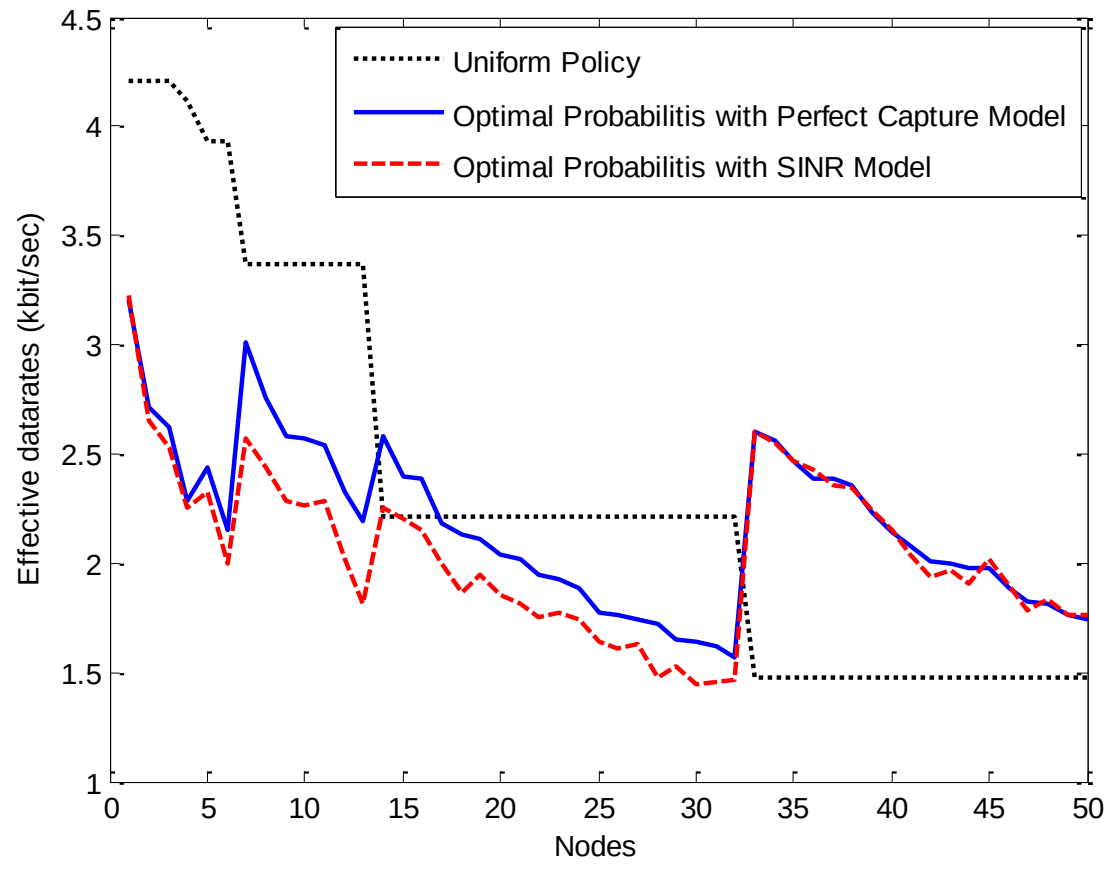

Figure 2. Effective data rates of nodes in a sample network

It is also of importance to examine performance of our algorithm in terms of lifetime (Figs 4). In order to do so, we define lifetime of the network as the time when 70% of nodes run out of energy. As we expected, the network will have greater lifetime for smaller values of $U_c$ (Fig4a). It is also apparent that our algorithm increases network lifetime in comparison with uniform policy (Fig4b).

## VII. Concluding Remarks

We presented a novel algorithm to enhance energy efficiency and guarantee fairness in random access networks. Based on simulation results, it has been verified that the proposed simple algorithm can reduce energy consumption of the network, enhance fairness among nodes, and increase network lifetime.

Although our algorithm has better lifetime characteristics than uniform policy, one valuable extension will be to optimize transmission parameters in order to directly maximize network lifetime. In order to do so, only step 4 of our algorithm will be changed and utility constrained lifetime elongation will be solved instead of UCEM. Another approach is to transform algorithm into a distributed structure. The results of section IV imply that any distributed algorithm with reasonable amount of messages will be suboptimal since power of all other nodes and number of nodes in each group should be known in order to achieve optimal values.

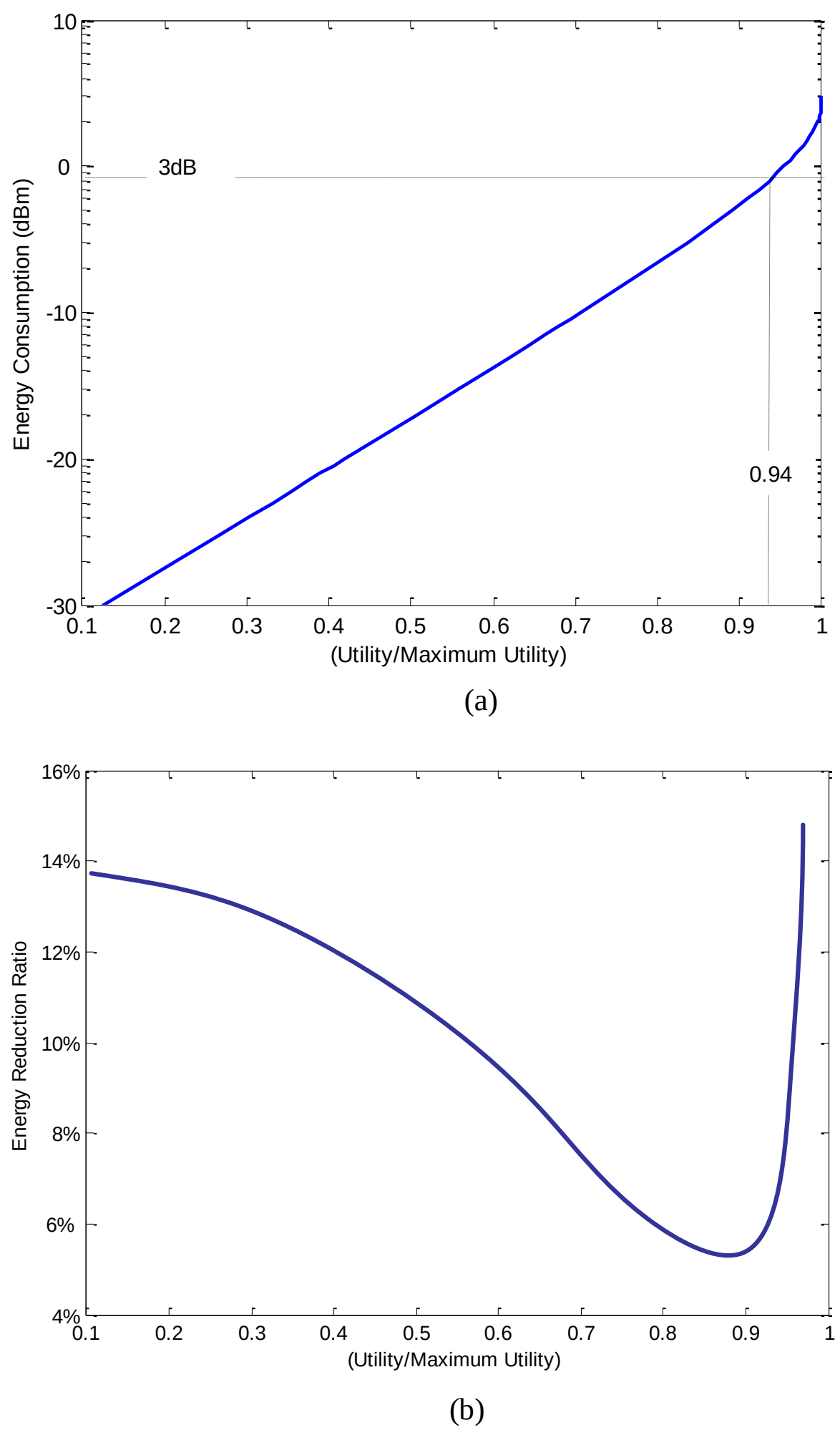

Figure 3. (a) Energy-utility tradeoff (b) Energy reduction of optimal transmission probabilities in comparison with uniform policy

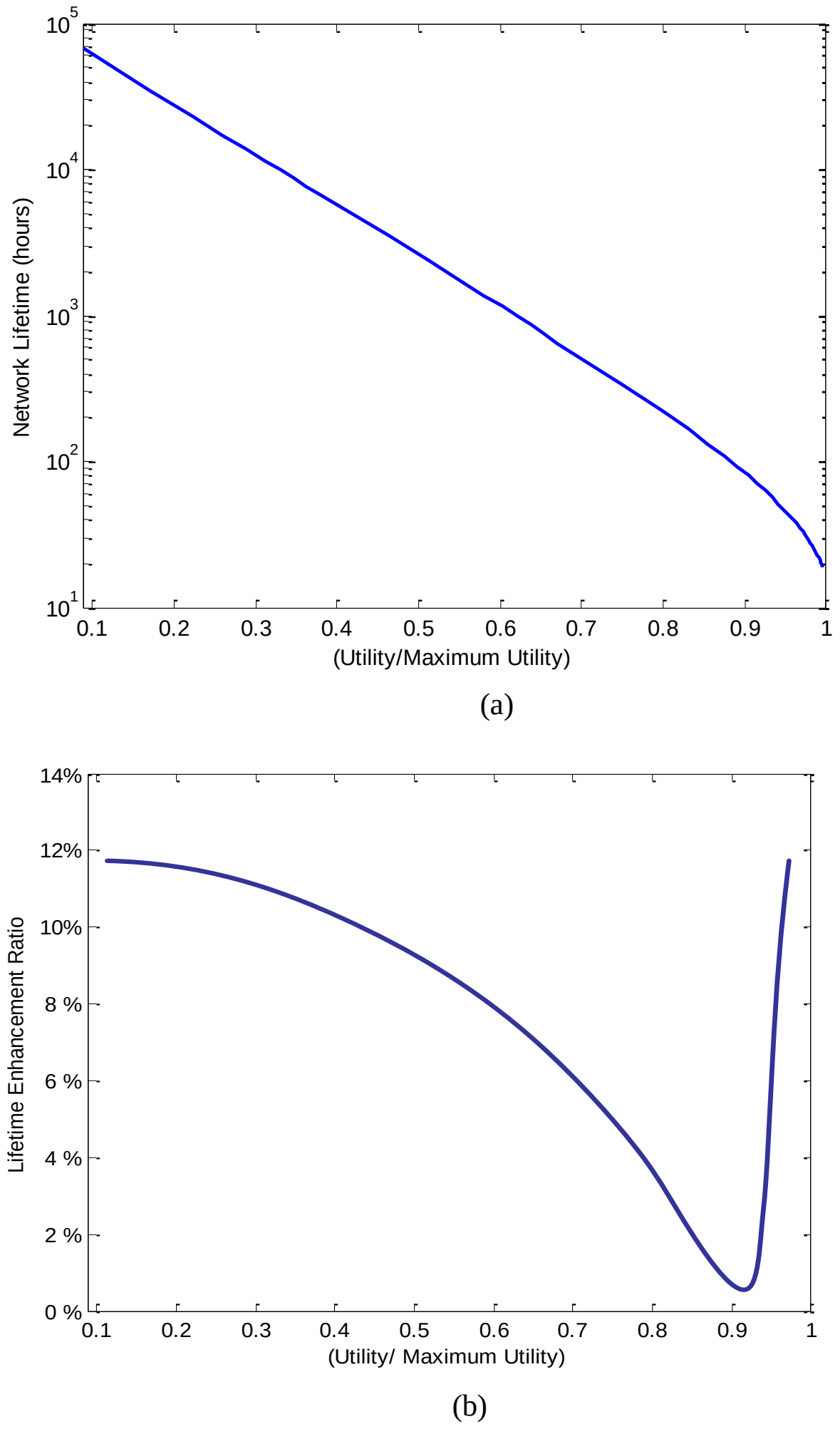

Figure 4. (a) Lifetime-utility tradeoff (b) lifetime enlongation of optimal transmission probabilities in comparison with uniform policy

## REFERENCES

[1] Norman Abramson, "The throughput of packet broadcasting channels" IEEE Trans. On Comm., Vol 25, No 1, pp 117-128, Jan. 1977.

[2] J. J Metzner, "On improving utilization in ALOHA networks", IEEE Transaction on Comm., Vol. 24, Issue 4, pp. 447-448, Apr. 1976.

[3] F. Berggren, J. Zander, "Throughput and energy consumption tradeoffs in pathgain based constrained power control in Aloha networks" IEEE Communication Letters, Vol 4, No 9, pp. 283-285, Sept. 2000.

[4] X. Qin, R. Berry, "Exploiting Multiuser Diversity for Medium Access Control In Wireless Networks", Proc. IEEE Infocom '03, pp 1084-1094, Apr. 2003

[5] T. Nandagopal, T. E. Kim, X. Gao, V. Bharghavan, "Achieving MAC layer fairness in wireless packet networks" In Proc. ACM/IEEE MobiCom '00, pp. 87-98, Dallas, USA, Oct. 1998

[6] A. Woo, D. Culler. "A transmission control scheme for media access in sensor networks". In Proc. MobiCom'01, Rome, Italy, pp. 221-235 july 2001.

[7] A.Silvester, T.-K.Liu, A.Polydoros. "Retransmission control and fairness issue in mobile slotted ALOHA networks with fading and nearfar effect". Mobile Networks and Applications, pp. 101-110, June 1997.

[8] K. Kar, S. Sarkar, and L. Tassiulas, "Achieving proportional fairness using local information in Aloha networks," IEEE Trans. on Automatic Control, vol. 49, no. 10, pp. 1858–1862, Oct. 2004.

[9] J. W. Lee, M. Chiang, R. A. Calderbank, "Jointly optimal congestion and contention control in wireless ad hoc networks", IEEE Comm. Letters, vol. 10, no. 3, pp. 216-218, March 2006.

[10] W. Li and H. Dai, "Optimal Throughput and Energy Efficiency for Wireless Sensor Networks: Multiple Access and Multi-packet Reception," Eurasip Journal on Wireless Communications and Networking, Vol. 5 , Issue 4, pp. 541-553, September 2005.

[11] A. Chockalingam, M. Zorzi, "Energy Efficiency of Media Access Protocols for Mobile Data Networks", IEEE Trans. on Communications, vol. 46, no. 11, pp. 541 – 553, November 1998.

[12] M. Medard, J. Huang, A. Goldsmith, S. Meyn, and TP Coleman, "Capacity of Time-slotted ALOHA Packetized Multiple-Access Systems over the AWGN Channel" IEEE Trans. On Wireless Comm. Vol. 3 No. 2, pp. 486- 499, March 2004.

[13] S. Boyd, L. Vandenberg, Convex Optimization, Cambridge University Press, 2004.

[14] R. Fletcher, Practical Methods of Optimization, Wiley, 1991